\documentclass[useAMS,usenatbib]{mn2e}
\usepackage{graphicx}
\usepackage{epstopdf}
\usepackage{amsmath}
\title[]{The Role of Outflows in dynamic of Advection Dominated Accretion Flows: a Self-Similar Solution}
\author[ F. Habibi, M. Ghasemnezhad]{
 F. Habibi $^{1}$\thanks{f\_habibi@birjand.ac.ir}, M. ghasemnezhad $^{2}$
\\
$^{1}$Department of Physics, Faculty of Sciences, University of Birjand, Birjand, Iran\\
$^{2}$Faculty of physics, Shahid Bahonar University of Kerman, Kerman, Iran}

\date{}
\begin{document}
\pagerange{\pageref{firstpage}--\pageref{lastpage}} \pubyear{2012}

\maketitle \label{firstpage}

\begin{abstract}
 The effects of outflow on the behavior of a viscous gaseous disc around a compact object in an advection-dominated state are examined in this paper. We suppose that the
flow is steady, axisymmetric, and rotating. Also, we focus on the model in which the mass, the angular momentum, and the energy
can be transported outward by outflow. Similar to the pioneering studies, we consider a power-law function for mass inflow rate as $\dot{M} \propto r^s$.
We assume that the power index $s$ is proportional to the dimensionless thickness $H/R$ of disc.
To analyze such a system, the hydrodynamic equations have extracted in cylindrical coordinates $(r,\varphi,z)$. Then, the flow equations were vertically integrated, and a set of self-similar solutions was got in the radial direction. 
Our solutions include three essential parameters: $\lambda$, $f$ and $\zeta$.
 The influence of the outflow on the dynamics of the disc is investigated by the $\lambda$ parameter.
The degree of advection of flow is shown by the advection parameter $f$. 
Also,  energy extraction from the disc by the outflow is showed by $\zeta$ parameter. 
Our findings demonstrate a significant correlation between the outflow parameters, flow advection parameter, and the temperature, thickness, and inflow-outflow rate of the disc.
In addition, we explored the influence of these parameters on the power index $s$, too. The results of our study demonstrate that enhancing the outflow parameter or flow advection degree increases power index $s$, while extracting more energy through outflow decreases index $s$.
\end{abstract}


\section{ Introduction}
Today, it is accepted that accretion discs surrounding compact objects exist in many cosmic phenomena, such as X-ray binaries (XRBs), quasars (QSOs), active galactic nuclei
(AGNs), and so on. For the research on the structure of such
systems, three main models are introduced: the standard
disc (Shakura \& Sunyaev 1973), the slim disc (Abramowicz et al. 1988), and the ADAF disc (Narayan \& Yi 1994, 1995a). The extract of heat produced by viscous dissipation
from the disc is one of the essential aspects that distinguish
these models. Radiation and advection are the two ways to transport this heating energy. The fundamental idea of the standard model is that the heat produced via viscosity
radiates out of the system, and so the disc becomes cold immediately (the flow temperature is far below virial temperature). 
This model works well for sources like regular luminous AGNs and the high/soft state of black hole binaries (Pringle 1981; Frank et al. 2002; Kato et al. 2008; Abramowicz \& Fragile 2013; Blaes 2014).
The slim discs are known as accretion systems with a high mass accretion rate and are geometrically slim. 
The disc in this state retains the viscous heating within the accreting fluid instead of radiating it locally.
Consequently, the radiation that is trapped moves inward alongside the accreting materials, leading to a reduction in radiative efficiency. This is what distinguishes these discs from standard discs (see, for instance, Kats 1977).
 The ultra-luminous X-ray sources, microquasars, luminous quasars, super-soft X-ray sources, and narrow-line Seyfert 1 galaxies belong to this branch of accretion discs (Fukue 2004; Kato et al. 2008).
ADAFs (or advection dominated accretion flows) are proposed as flows with a low mass
accretion rate and geometrically thick. In this scenario, advective cooling mainly counterbalances viscous heating while radiation is not effective.
 Then, the disc becomes hot (the gas temperature is close to the virial temperature, i.e., $10^4–10^7 K$) and optically thin.
The suggested model attempts to account for various sources including the supermassive black hole in the galactic center (Sgr A*), low-luminosity AGNs, and black hole XRBs in the hard state  (e.g., Narayan \& McClintock 2008; Yuan \& Narayan 2014).

According to certain theoretical works, the mass accretion rate is assumed to stay constant throughout the process of accretion. Nevertheless, some numerical simulations suggest that the inflow rate decreases as it moves inward (e.g., Stone et al. 1999; Yuan et al. 2012), which can be explained physically by the outflow of mass. Recent observations have also verified the presence of outflows in the thin discs (e.g., King \& Pounds 2015; Díaz Trigo \& Boirin 2016; Homan et al. 2016), slim discs (e.g., Gladstone et al. 2009; Middleton et al. 2011; Du et al. 2015), and ADAFs (e.g., Wang et al. 2013; Cheung et al. 2016; Homan et al. 2016; Ma et al. 2019; MuñozDarias et al. 2019). Mass, angular momentum, and energy can be removed from the accretion disc by an outflow, resulting in a modified structure of the accreting flow. Hence, outflows have garnered significant attention from researchers and expanded into various studies. For instance, Narayan \& Yi (1994, 1995a) argued that a necessary condition for outflows is that the Bernoulli
parameter becomes positive. Blandford and Begelman (1999) successfully obtained a global inflow-outflow solution to the adiabatic flow that was quite innovative.
They postulated that the accretion rate follows a power-law function with radial dependence as: $\dot{M} \propto r^s$, where $s$ ranges from $0$ to $1$.
Xie and Yuan (2008) used this connection to show how the outflows impact the structure of the disc. The accretion rate's power law function has received additional support from subsequent numerical simulations (e.g. Yuan et al . 2012a; 2012b; Bu et al. 2013; Yang et al. 2014; Bu \& Gan 2018). However, the range of changes of power-law index $s$ was determined from $0.5$ to $1$ in some works (e.g., Stone et al. 1999; Ohsuga et al. 2005; Narayan et al. 2012; Yuan et al. 2012b; Bu \&
Gan 2018). Regarding $Sgr A^*$, a relatively low amount for the index $s$ is mentioned,
such as $s = 0.25$ (Quataert \& Narayan 1999), $0.27$ (Yuan et al. 2003), and $0.37$ (Ma et al. 2019).
It appears that some disc characteristics may influence the power index of the accretion rate's dependence on the radius.
 On the flip side, numerical simulations have indicated that outflows in slim discs and ADAFs are more potent than in standard discs (e.g., Ohsuga \& Mineshige 2011, 2014 ). It can be physically understood as follows.
The cooling of radiation in standard discs effectively lowers the disc temperature, resulting in a negative flow Bernoulli parameter. Hence, standard discs have the capacity to generate only relatively weak outflows. On the other hand, in cases of slim discs, although the disc temperature is only slightly elevated compared to standard discs, the large number of trapped photons intensifies radiation pressure, giving rise to strong outflows. Additionally, in the case of ADAFs, the disc's temperature increases significantly due to energy advection, leading to a positive Bernoulli parameter and strong outflows (e.g., Narayan \& Yi 1994; Narayan et al. 1997b). In brief, the physics of advection, whether involving photons or gas, aids in producing robust outflows. The disc's advection strength is accurately captured by the dimensionless thickness $H/R$, where $H/R \ll 1$ for standard discs and $H/R \leq 1$ for slim discs and ADAFs. According to this, Wu et al. (2022; here after WU22) reasoned that the power law index $s$ should be proportional to dimensionless thickness $H/R$ of disc.
Subsequently, they acquired the thermal equilibrium solutions for the accretion flow around a black hole and conducted a comparison between the cooling rate of the outflow and advection.

Taking this background into account, our current aim is to investigate the potential effects of outflow on the dynamics and structure of an advection dominated accretion flow.
Like the aforementioned studies, we also include a power law function for the mass inflow rate. Next, we will analyze the disc's radial structure using the self-similarity formulation. Currently, semi-analytical approaches serve as useful tools for comparing theoretical models to observational data. Although some works have been made to solve hydrodynamic
equations in accretion discs with outflow (e.g., Zeraatgari et al 2016; Ghasemnezhad \& Abbassi 2017; Ghoreysh 2020;
Ghoreyshi \& Shadmehri 2020), all of which have considered arbitrary values for power index
$s$. In this paper, we are going to gain a more realistic picture of this problem by
including a non-constant power law index that depends on the disc characteristics. 
As stated above, we can consider power index $s$ proportional to
dimensionless thickness $H/R$ of the disc. Such an assumption is appropriate for different models of accretion with outflow.
The paper is organized as follows:  In Section 2; we derive the basic equations governing the behavior of the system. Using a self similar technique in the radial direction, we simplify theses equations in Section 3. We describe the numerical results of our model with detail in Section 4. We then abridge our main findings in the final part.

\section{BASIC EQUATION}
As mentioned in the introduction, we are going to analyze the structure and dynamic of an accretion flow in which outflows play a serious role. For this purpose, we consider a hot plasma disc around a compact object with mass $M$.
We assume that disc is stationary ($\partial/\partial t= 0$), axisymmetric ($\partial/\partial \varphi= 0$) and geometrically thick ($H/R \leq 1$). We apply the cylindrical coordinate system ($R$, $\varphi$, $z$) to formulate the system equations and consider its origin on the central
object. To pass relativistic effects, we employ the Newtonian potential. We also ignore the
self-gravity of the disc. Furthermore, we suppose that dominant component of viscose stress
tensor is $\varphi$-component and viscosity of rotating gas follow $\alpha$-prescription.
Moreover, we consider height-integrated set of equations which allow us to
describe all our physical variables as function
 of the cylindrical radius $R$ only . Therefore, the fundamental equations
of our system can be written as follows:\\
 The continuity equation is: 
\begin{equation}
 -\frac{d\dot{M}}{dR} +\frac{d\dot{M}_w}{dR}=0,\label{1}
\end{equation}
  where $\dot{M}$ is the mass accretion rate and $\dot{M _w}$ is the mass loss rate. Following the pioneering work done by Blandford \& Begelman (1999, 2004), we consider the radial dependence of the mass inflow rate as:
 \begin{equation}
  \dot{M}=\dot{M}_{out}\big(\frac{R}{R_{out}}\big)^s.\label{mdot1}
 \end{equation}
 wherein $\dot{M}_{out}$ is the mass accretion rate at the outer boundary $R _{out}$. The parameter $s$ determines how the accretion rate is modified. For a disc with outflow, $s$ is considered more than zero $s>0$,
while in the absence of outflow, $s = 0$ is used (Fukue 2004). In most previous analytical
works on ADAFs and slim discs, the power law index $s$ were assumed to be constant (Quataert \& Narayan 1999; Beckert 2000; Fukue 2004;
Ghasemnezhad \& Abbassi 2017). Following Wu22, we adopt that the power-law index $s$ is proportional to the dimensionless thickness $H/R$ of the
disc, i.e ., $s =\lambda(H/R)$, in which $\lambda$ is a constant coefficient.
In our opinion, this idea can present a more realistic picture for different accretion models.
  On the other hand, the mass loss rate is introduced as follows:\\
\begin{equation}
\dot{M}_w(R)=\int {4 \pi r \dot{m}_w(r)}dr.\label{mdot2}
 \end{equation}
where $\dot{m}_w$ is the mass loss rate per unit area from each disc face. Utilizing Equations (\ref{1})–(\ref{mdot2}), we arrive to this relation for $\dot{m}_w$: 
\begin{equation}
\dot{m}_w=\lambda \big(\frac{H}{R}\big) \dot{m}_{out}\bigg(\frac{R}{R_{out}}\bigg)^{\lambda(H/R)-2}.\label{12}
\end{equation}
 wherein $\dot{m}_{out}=\dot{M}_{out}/(4\pi R_{out}^2)$ is the mass loss rate per unit area at the outer boundary. One can see that for a bigger coefficient of $\lambda$, the mass loss rate increases. Thus, $\lambda$ parameter can be called as outflow parameter.
 
 The equations of motion in the radial and azimuthal directions can be respectively written as follows: 
\begin{equation}
 V \frac{dV}{dR}+ R(\Omega_K^2 -\Omega^2)+\frac{1}{\Sigma}\frac{d}{dR}\big(\Sigma c _s^ 2\big)=0,\label{2}
\end{equation}
and
\begin{equation}
V \frac{d}{dR}(R^2\Omega)-\frac{1}{\Sigma R}\frac{d}{dR}\bigg(\nu \Sigma R^3\frac{d\Omega}{dR}\bigg)+\frac{l^2 \Omega R}{ 2 \pi \Sigma}\frac{d\dot{M}_w}{dR} =0,\label{3}
\end{equation}
In the equation (\ref{2}), $c_s^2$ is the square of the sound velocity.
This speed that is proportional to system temperature define as the ratio of gas pressure to density: $c_s^2=p_{gas}/\rho$. 
 Also, $\nu$ in the equation (\ref{3}) is used as the
kinematic constant of viscosity, formalized by Shakura \& Sunyaev (1973 ) as
 \begin{equation}
  \nu=\alpha c _s H, 
  \end{equation}
wherein $\alpha$ is a constant less than unity and is known as the viscous parameter. Moreover, the third term in
equation (\ref{3}) demonstrates the angular momentum removed by the outflowing material.
The $l$ parameter is defined as the length of the rotational lever arm and can identify the outflow type.
The value of $l = 0$ corresponds to a non-rotating outflow in which the angular momentum can not be extracted by the outflowing material. The
case with $l = 1$ shows a type of outflow in which outflowing material carries away the specific angular momentum $R^2 \Omega$. The cases with $l < 1$ (or $> 1$) represent the outflows in materials that extract less (or more) angular momentum from the disc surface( Knigge 1999 ). In the present paper, $l$ is
assumed to equal $1$.

 In the vertical direction, the gravity force should be balanced with the pressure gradient.
Thus, the z-component of the motion equation yields as: 
\begin{equation}
 \rho \frac{\partial \Phi}{\partial z} + \frac{\partial p _{gas}}{\partial z} =0,
 \end{equation}
Here, $\Phi=- GM/(R^2 +z^2)^{1/2}$ is the Newtonian potential
in cylindrical coordinates. With the aid of some approximations, the hydrostatic balance leads to a relation between $H$ and $c_s$ in the following form: 
 \begin{equation}
  \frac{GM}{R^3}H^2 =\Omega_k^ 2 H^2=c_s^2,\label{4}
 \end{equation}
  wherein $\Omega_k=\sqrt{GM/R^3}$ is the Keplerian angular velocity .
  
Finally, the conservation of energy can be used by considering the cooling and heating processes
in the system. It is possible to extract the energy produced by viscous dissipation, $Q_{vis}$, from the disc through radiation and outflowing material, as well as transport it to the central object through advection. 
Therefore, the energy equation is introduced as:
 \begin{equation}
  Q_{vis}-Q_{rad}=Q_{adv}+Q_w,\label{energy}
 \end{equation}
  wherein $Q_{rad}$, $Q_w$ and $Q_{adv}$ are radiative cooling, cooling due to outflow and advective cooling, respectively. To determine the
advection degree of flow, we introduce the advection parameter $f$ as $f = 1- \frac{Q_{rad}}{Q_{vis}}$ (Narayan \& Yi 1995). So, we can substitute $f Q_{vis}$ with $Q_{vis}-Q_{rad}$ on the
 left-hand side of the equation (\ref{energy}). In general, advection parameter $f$ is affected by the details of heating and cooling processes
of disc ( Watarai 2006, 2007, Sinha et al. 2009). However, all the results we present here are derived from the straightforward assumption suggested by Narayan \& Yi (1994), that is, $f=constant$. Therefore, in the case of $f=1$, also known as advection-dominated, we can disregard radiative cooling. In contrast, when $f=0$ and the system is radiation-dominated, advection cooling is negligible. Now, with the assistance of definitions $Q_{vis}$, $Q_{adv}$ and $Q_w$ as
follows: 
\begin{equation}
 Q _{vis}=\nu \rho R^2(\frac{\partial \Omega}{\partial R})^ 2,
 \end{equation}
 \begin{equation}
  Q_{adv}=\Sigma V \bigg(\frac{1}{\gamma-1}\frac{dc_s^ 2}{dR}- \frac{c_s^2 }{R} \frac{d\rho}{dR}\bigg). 
\end{equation}
\begin{equation}
 Q_{w}=\frac{1}{2} \zeta \dot{m_w} R^2 \Omega_k^2,
 \end{equation}
 we can rewrite the energy equation:
\begin{equation}
 \Sigma V \bigg(\frac{1}{\gamma-1}\frac{dc_s^ 2}{dR}- \frac{c_s^2}{R}\frac{d\rho}{dR}\bigg)=f \nu \rho R^2(\frac{\partial \Omega}{\partial R})^2-\frac{1}{2}\zeta \dot{m_w} R^2 \Omega_k^2.\label{5}
\end{equation}
The dimensionless parameter $\zeta$ is used as an outflow energy parameter in the equations above.
To account for energy loss from the disc via the outflows, we consider it as an arbitrary parameter in our model, with larger $\zeta$ values indicating greater energy loss (Knigge 1999).

In conclusion, the behavior of system can be described by equations (\ref{12}), (\ref{2}), (\ref{3}), (\ref{4}) and (\ref{5}). By solving these equations, one can obtain the dynamical properties of the disc. To solve these equations, we will employ the similarity method in the upcoming section.

  \section{Self-Similar Solutions}
 The similarity technique is based on dimensional analysis and is widely applied in astrophysical fluid mechanics (Kato et al. 2008, Narayan \& Yi 1994). Generally, similarity methods can be used in two problem types: radial similarity for steady problems and temporal similarity for unsteady problems.
In the radial self-similar approach, which is used in this paper, we can search for answers that the radial alteration of physical quantities at any distance from the center of the system is similar to the others.
Such answers can be obtained by power functions from radial distance. Hence, when considering a constant radial distance $R_{out}$, the physical quantities can be described by a power-law expression of $R/R_{out}$. The exponents are determined by satisfying the fundamental equations in a self-consistent manner. The general behavior of an accretion flow cannot be explained by a radial self-similar solution, as it does not consider any boundary conditions. These solutions are applicable only in regions distant from the inner and outer boundaries of the disc.
Consequently, by disregarding problem boundaries, similarity solutions can accurately depict disc dynamics in an intermediate region and aid in comprehending the underlying physical processes.
The validity of similarity solutions in distant regions from boundaries has also been supported by numerical simulations (Stone, Pringle \& Begelman 1999, De Villiers et al 2005, Beckwith et al. 2008, Yuan et al 2012).
Now, we introduce expressions for the radial velocity $V$, angular velocity $\Omega$, sound speed $c_s$ and the surface density $\Sigma$ as follows:
 \begin{equation}
  V=-v_0 \sqrt{\frac{GM}{R_{out}}} \bigg(\frac{R}{R_{out}}\bigg)^{-\frac{1}{2}}, 
\end{equation}
\begin{equation}
  \Omega=\omega_{0} \sqrt{\frac{GM}{R_{out}^3}}\bigg(\frac{R}{R_{out}}\bigg)^{-\frac{3}{2}},
   \end{equation}
 \begin{equation}
  c_s= c_0 \sqrt{\frac{GM}{R_{out}}} \bigg(\frac{R}{R_{out}}\bigg)^{-\frac{1}{2}},\label{cs1}
 \end{equation}
  \begin{equation}
   \Sigma= s_0 \Sigma_{out} \bigg(\frac{R}{R_{out}}\bigg)^{s-\frac{1}{2}},\label{sigma1}
 \end{equation}
 wherein $v_0$, $\omega_0$, $c_0$ and $s_0$ are dimensionless coefficients and will be determined later. 
\begin{figure*}
 \centering
 \includegraphics[width=180mm]{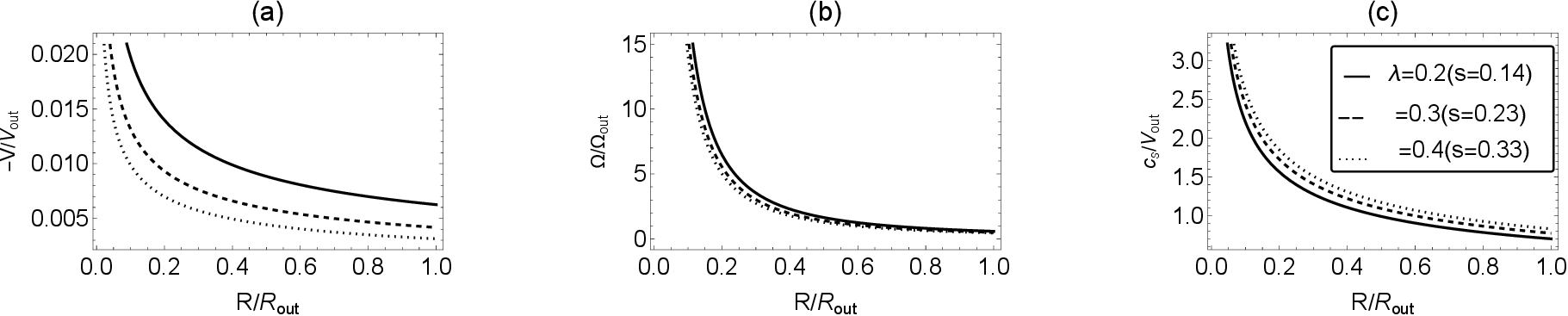}
\includegraphics[width=180mm]{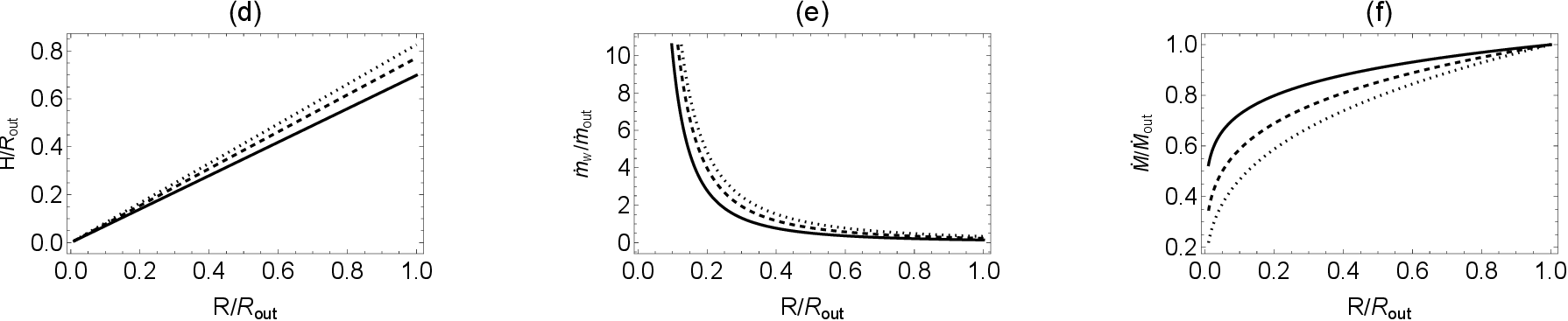}
 \caption{The physical quantities of disc as a function of radial distance $R$ and several values of
$\lambda$. The values of $\lambda$ are given in the legend. We also show the values of corresponding to $s$ for each value of $\lambda$.}\label{fig1}
 \end{figure*}
\begin{figure*}
 \centering 
 \includegraphics[width=180mm]{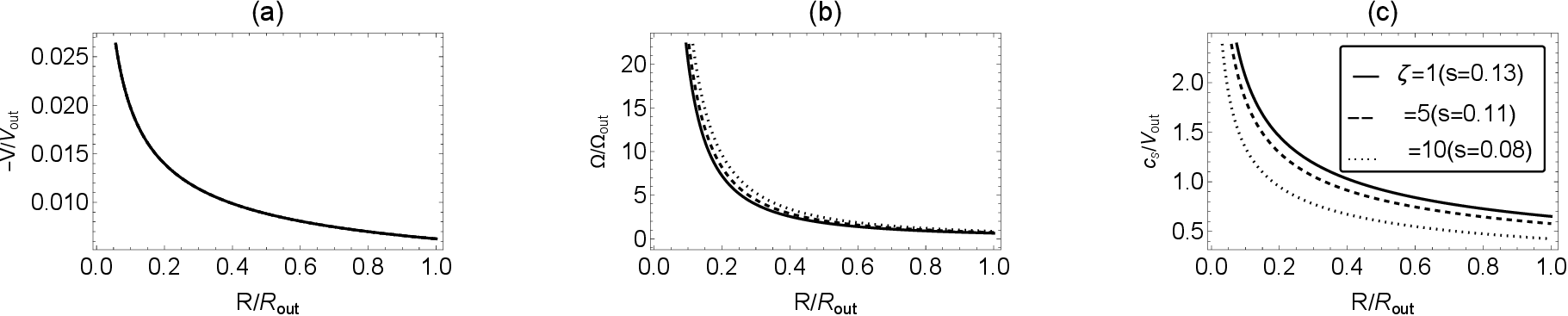}
\includegraphics[width=180mm]{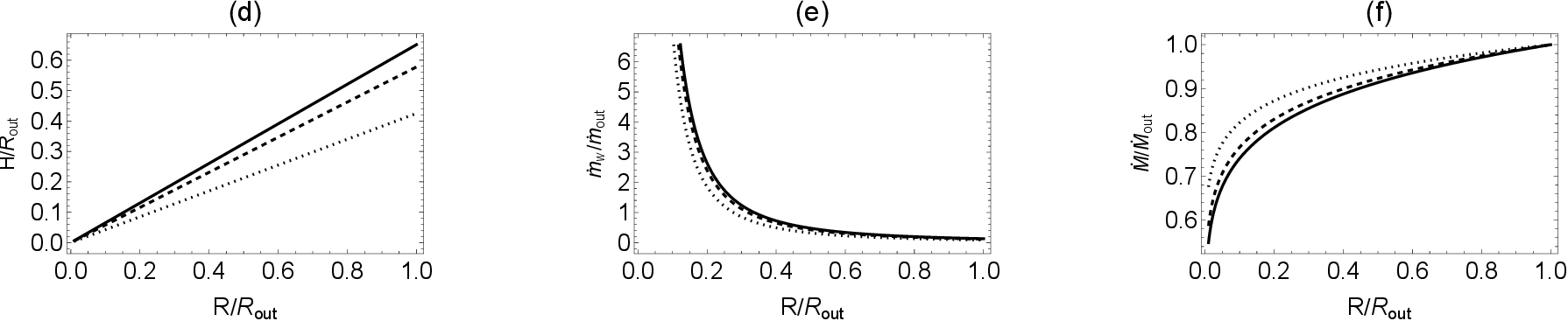}
 \caption{Similar to Figure \ref{fig1}, but for
different values of $\zeta$ (as labelled).}\label{fig2}
 \end{figure*}
 \begin{figure*}
 \centering
 \includegraphics[width=180mm]{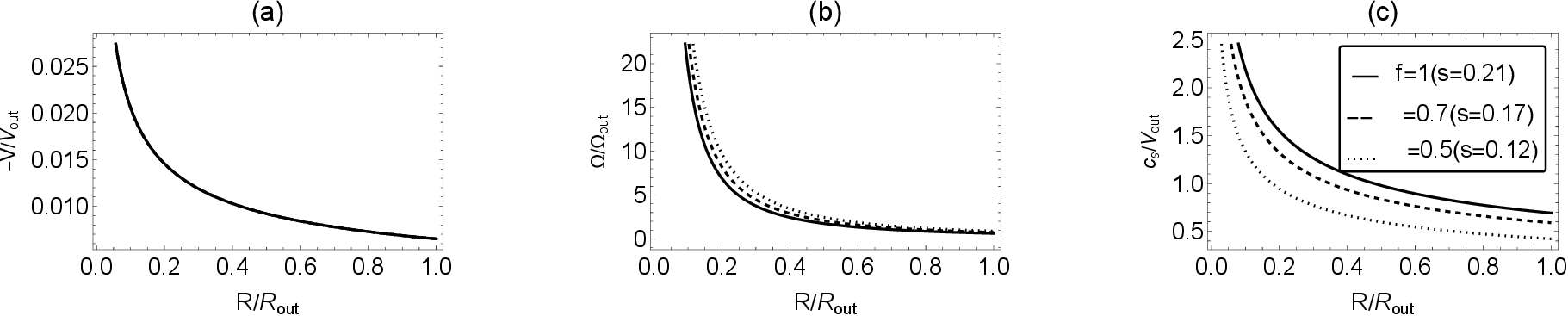}
\includegraphics[width=180mm]{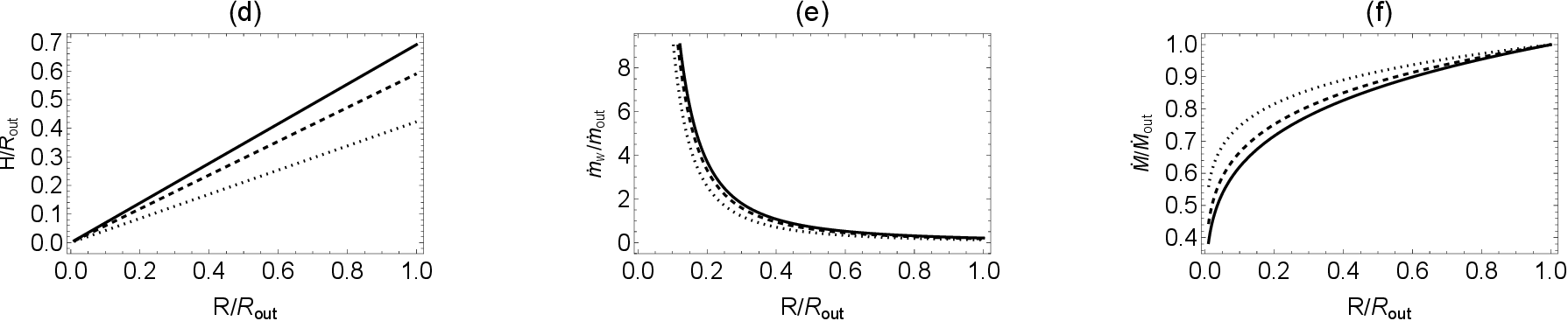}
 \caption{Similar to Figure \ref{fig1}, but for different values of $f$ (as labelled).}\label{fig3}
 \end{figure*}
  \begin{figure*}
 \centering
 \includegraphics[width=180mm]{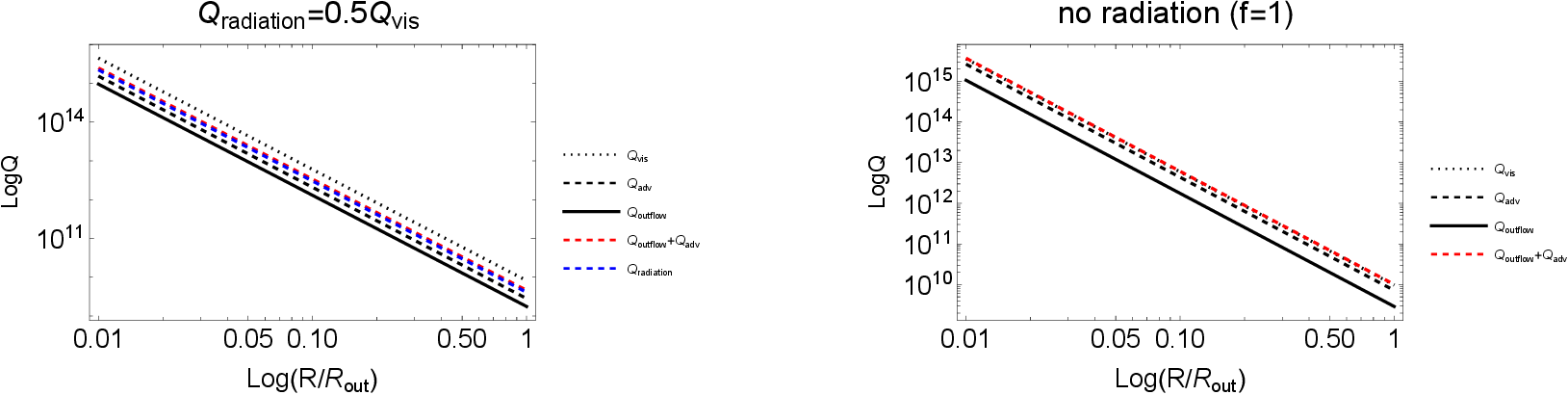}
 \caption{Similar to Figure \ref{fig1}, but for different values of $f$ (as labelled).}\label{fig3}
 \end{figure*}
 Also, $\Sigma_{out}$ is used in order to write equations in a non-dimensional form. Replacing the above self-similar
transformations in equations of the system, we obtain the following set of equations, which should be solved to
determine the coefficients of dimensionless: 
\begin{equation}
s_0=\frac{\dot{m_0}}{v_0},\label{s1}
 \end{equation}
  \begin{equation}
   v_0^2 - 2 + c_0^2(3 - 2\lambda c_0) + 2 \omega_0^2=0.
 \end{equation}
  \begin{equation}
  9 f \alpha \omega_0^2 c_0^2-v_0(4\lambda c_0^2-6\epsilon c_0+\zeta\lambda)=0, 
\end{equation}
 \begin{equation}
\big(A+2\sqrt{2B}\big)^{1/3}\big(1+6\lambda c_0 -8\lambda v_0+\alpha +\frac{(A+2\sqrt{2B})^{1/3}}{\alpha}\big)=0.\label{s4}
 \end{equation}
  \begin{equation}
   H/R=c_0,\label{h}
    \end{equation}
  where $\dot{m_0}=\frac{\dot{M_0}}{2 \pi \sqrt{G M} r_{out} \Sigma_{out}}$ is the non dimensional mass accretion rate and 
  \begin{equation}
   A=\alpha^3 - 12v_0 \alpha^2 \lambda(1+3\lambda). 
   \end{equation}
\begin{equation}
 B=v_0\alpha^3 \lambda^2 \big[-6 v_0 \alpha +64 \lambda v_0^2 +108 \alpha \lambda v_0 -9\alpha(\alpha -18 \lambda^2 v_0)\big], 
\end{equation}
 \begin{equation}
  \epsilon=\frac{\gamma-5/3}{\gamma-1}.
 \end{equation}
  Also, for the density, we have:
   \begin{equation}
\rho=\frac{1}{2}\frac{s_0 \Sigma_{out}}{c_0 R_{out}}\bigg(\frac{R}{R_{out}}\bigg)^{-3/2},
\end{equation}
 The unknown coefficients of $v_0$, $c _0$ and $\omega_ 0$ can be specified by solving the equations (\ref{s1})-(\ref{s4}) numerically. To investigate the
disc properties physically, only real roots must be adopted. Also, our results are dependent on free parameters such as the advection parameter $f$, the radio of the specific heats $\gamma$, the outflow parameter $\lambda$, the standard viscose parameter $\alpha$, and the outflow energy parameter $\zeta$. Among these
parameters, the outflow parameter $\lambda$, the outflow energy parameter $\zeta$, and the advection parameter $f$ are the essential parameters in our system. Hence, in next section, we illustrate the behavior of disc variables for some values of $\lambda$, $\zeta$ and $f$. We fix $\gamma=4/3$ and $\alpha=0.01$.

 \section{ Numerical results}
 
In this section, we present the numerical results of our model with focusing on the effects of outflow and advection degree of flow. To gain a quantitative understanding of the results at different radii, we plot disc variables as a function of the radial distance $R$.. Figure \ref{1} shows the variations of the physical quantities of the disc for different amounts of
the outflow parameter, $\lambda= 0.2$ (solid line), $0.3$ (dashed line), and 0.4 (dotted line).
Then, the power law index $s$ has been found to be $0.14, 0.23$, and $0.33$, respectively (see equation 3).
 These values are consistent with the fitting results of the Sgr A* observations (e.g., Quataert \& Narayan 1999; Yuan et al. 2003; Ma et al. 2019 ).
By considering a fully advection flow (i. e., $f=1$), the behaviour of the radial velocity, angular speed, sound velocity (temperature) and thickness of disc are showed in panels (a)–(d) of Figure \ref{1}, respectively. Moreover, we plot the mass loss rate $\dot{m_w}$ and mass accretion rate $\dot{M}$ in panels (e) and (f). As these plots clearly show, by adding the $\lambda$ parameter the mass outflow rate rises (panel e), while the mass inflow rate decreases (panel f). Additionally,
as we expect, the more substantial outflows occur in the thicker and hotter discs (panels c and d). However, the enhance of disc thickness is vital in outer regions. Furthermore, as seen in
panels (a) and (b), a thicker disc which has stronger outflow rotates slower and less matter can also flow toward the central object.

 The effect of outflow energy parameter $\zeta$ on the physical quantities of the disc for
the case of without radiative cooling ($f=1$) is shown in Figure (\ref{fig2}). According to
Panels (a) and (b), the parameter $\zeta$ cannot significantly affect the radial velocity, but
removing energy by outflow enhances the rotational velocity. Looking at panels (c) and (d)
reveal that with more energy extraction from the disc by outflow, temperature and vertical
thickness decrease. This result is also obtained by previous numerical calculations (e.g., Faghei \& Mollatayefeh 2012; Ghoreyshi \& Shadmehri 2020). Moreover, one can see that the effect of $\zeta$ parameter on disc thickness becomes weaker as we approach central object. Additionally, the mass loss rate $\dot{m_w}$ has a
decreasing trend concerning $\zeta$ in a constant radius. The decline of outflow rate is along with the rise of inflow rate, as can be seen in panel (e) from Figure (\ref{fig2}).
 Here, it is worth mentioning that the power law index $s$ is sensitive to the variations of the
parameter $\zeta$ and reduces with more energy extraction. In other words, in the present
model, $s$ and $\zeta$ vary inversely with each other. This dependency is created through
dimensionless thickness $H/R$, implicitly. 

Figure \ref{fig3} is plotted similarly to Fig. \ref{fig2} but for three values of the advection parameter: $f= 1$ (solid line), $f=0 .8$ (dashed line ), and $f=0.6$ (dotted line). We then obtain variations of $s$ as $0.21$, $0.19$, and $0.15$, respectively. As a result, there is a direct relationship between the power law index $s$ and the degree of advection of flow $f$. In addition, panel (a) displays that the change of radial velocity
due to the advection parameter $f$ is not significant. However, once $f$ takes higher values,
the angular speed slows down, as seen from panel (b) of figure \ref{fig3}. Moreover, with
reducing degree of advection, the temperature and thickness of disc decrease. This is
reasonable because when $f$ become smaller the energy contribution released via radiation
increases and disc becomes colder and thinner.
Furthermore, when the level of advection is diminished, the mass inflow rate displays an ascending behaviour while the mass outflow rate shows a descending behaviour (see
panels e and f).

  \section{conclusion}
This paper examines how the outflow affects the dynamics and structure of an advection-dominated accretion flow around a central object. Describing the macroscopic behavior of this system is possible through hydrodynamic equations. We made some approximations in the main equations of the system for the sake of simplicity: (1) The gravity of the central mass was determined using the Newtonian potential, (2) The disc's self-gravity and general relativistic effects were not taken into account, (3) The flow was supposed to be steady and axisymmetric, (4) It was assumed that the outflow had the ability to remove mass, angular momentum, and energy from the disc, (5) Similar to the pioneering works, a power-law function for the accretion rate was considered as $\dot{M} \propto r^s $, (6) Following the methodology of Wu22, we presumed the power index $s$ be function of
dimensionless thickness $H/R$. Subsequently, the vertical integration of the conservation equations for mass, momentum, and energy was performed in cylindrical coordinates. By utilizing a self-similar approach, we successfully solved these equations. The role of outflow in the dynamics of flow was shown by
the outflow parameter $\lambda$ and the energy parameter of outflow $\zeta$. We also
employed the advection parameter $f$ between $0$ (radiative cooling locally) and
$1$ (no radiative cooling).

 Our results showed that the radial dependence of mass accretion rate, i.e., the value of
power index $s$, is a function of three essential parameters of model. Namely, the
value of $s$ reduces with increasing the energy parameter $\zeta$ and rises when
$\lambda$ and $f$ become larger. Moreover, our self-similar solutions indicated that an
increase in the outflow parameter $\lambda$ lead to more thickness and temperature of disc.
This means that outflows are stronger in hotter and thicker discs. The increase of disc
thickness decreases the rotational velocity and less matter can also flow toward central
object. However, as more energy is removed from the disc by matter, the disc will become
colder and thinner. This can lead to a high mass accretion rate. Additionally, when we increase the
energy contribution released via radiation by a reduction in $f$ parameter, disc rotates faster
and becomes colder and thinner. These results are physically well understood, and
are in good agreement with previous findings.

Radially one-dimensional self-similar solutions were employed here. Despite
simplifications considered in this model, our results provide a well understanding from the dynamics of an advection
dominated accretion flow with outflow. Nonetheless, to achieve a more accurate understanding of the outflow, studying these systems in two dimensions is significantly better. In future works, this problem can be pursued.
Furthermore, in a practical model, the advection parameter $f$ varies with position and time; other researchers can consider this.
Moreover, in the present study, parameter $l$ was used to represent the angular momentum extracted by the outflow, with a fixed value of $1$.
The disc structure can be significantly affected by outflows that extract more angular momentum, specifically those with $l > 1$, like the centrifugally driven magnetohydrodynamic outflows mentioned by Blandford \& Payne in 1982. This topic is open for investigation in other studies.

\section{DATA AVAILABILITY STATEMENT}
No new data were generated or analysed in support of this research.

\end{document}